\documentclass[aps, twocolumn, superscriptaddress, floatfix]{revtex4-2} % showkeys, showpacs

\usepackage{graphicx}
\usepackage[utf8]{inputenc}
\usepackage{amssymb, amsmath, commath, braket, physics, soul}
\usepackage{hyperref}
\hypersetup{
colorlinks=true,
linkcolor=red,
filecolor=blue,
urlcolor=magenta,
}
\usepackage{placeins, standalone}
\newcommand{\Eth}{\ensuremath{E_{\textrm{Th}}}}	%	Thouless energy
\newcommand{\Ecal}{\ensuremath{\mathcal{E}}}	%	Unfolded energy

\newcommand{\fbsli}[2]{\ensuremath{\textrm{I}_{#1}\del{#2}}}
\newcommand{\fbslj}[2]{\ensuremath{\textrm{J}_{#1}\del{#2}}}

\newcommand\mean[1]{\ensuremath{\left\langle #1 \right\rangle}}

\newcommand{\sff}[1]{\ensuremath{\mathcal{K}\del{#1}}}
\newcommand\sffbar{ \ensuremath{ \overline{\mathcal{K}} } }

\newcommand\tdip{\ensuremath{t_\mathrm{dip}}}
\newcommand\tH{\ensuremath{t_\mathrm{H}}}
\newcommand\tR{\ensuremath{t_\mathrm{R}}}
\newcommand\tZ{\ensuremath{t_\mathrm{Zeno}}}

\begin{document}
\title{Spectral form factor and energy correlations in banded random matrices}
\author{Adway Kumar Das}
%\affiliation{Department of Physics, University of Connecticut, Storrs, Connecticut 06269, USA}
\affiliation{Department of Physical Sciences, Indian Institute of Science Education and Research Kolkata, Mohanpur 741246, India}
\author{Anandamohan Ghosh}
\affiliation{Department of Physical Sciences, Indian Institute of Science Education and Research Kolkata, Mohanpur 741246, India}
\author{Lea F. Santos}
\affiliation{Department of Physics, University of Connecticut, Storrs, Connecticut 06269, USA}

%=================================
\begin{abstract}
Banded random matrices were introduced as a more realistic alternative to full random matrices for describing the spectral statistics of heavy nuclei. Initially considered by Wigner, they have since become a paradigmatic model for investigating level statistics and the localization-delocalization transition in disordered quantum systems. In this work, we demonstrate that, despite the absence of short-range energy correlations, weak long-range energy correlations persist in the nonergodic phase of banded random matrices. This result is supported by our numerical and analytical studies of quantities that probe both short- and long-range energy correlations, namely, the spectral form factor, level number variance, and power spectrum. We derive the timescales for the onset of spectral correlations (ramp) and for the saturation (plateau) of the spectral form factor. Unexpectedly, we find that in the nonergodic phase, these timescales decrease as the bandwidth of the matrices is reduced. We also show that the high-frequency behavior of the power spectrum of energy fluctuations can distinguish between the nonergodic and ergodic phases of the banded random matrices.
\end{abstract}

%============================================================
% 	PACS, the Physics and Astronomy Classification Scheme.
% 	05.45.Mt: Quantum chaos; semiclassical methods
%	02.10.Yn: Matrix Theory
%	89.75.Da: Systems obeying scaling laws
%\pacs{05.45.Mt, 02.10.Yn, 89.75.Da}
%--------------------------------------------------------
%	Use showkeys class option if keyword display desired
%\keywords{Spectral form factor, correlation hole, power-spectrum of noise, banded random matrix}

\maketitle
%=================================
\section{Introduction}
%=================================

Matrices filled with random numbers were originally studied by mathematicians in the context of probability theory and statistics. Their significance grew in the 1950s with Wigner's works, where they were used to model the spectral properties of heavy nuclei~\cite{Wigner1958, Dyson1962, MehtaBook}. These studies were soon extended to other complex systems, including atoms and molecules~\cite{Guhr1998}. The main advantage of full random matrices is their analytical tractability. However, they are not physically realistic, as they imply all-to-all couplings and simultaneous interactions among all the particles. To better reflect the fact that the interactions in physical systems are local and decay with distance, Wigner introduced banded random matrices (BRMs)~\cite{Wigner1955}.

The entries of BRMs are non-zero only near the main diagonal~\citep{Wigner1955, Brody1981, Casati1996, Fyodorov1996}. The band of nonzero elements may be associated with a preferential basis~\cite{Shepelyansky1994,Oppen1996} and the matrices can also exhibit sparsity~\cite{Prosen1993,Fyodorov1996} or correlated entries~\cite{Janssen2000}. By increasing the bandwidth, BRMs can model the transition from integrability, characterized by Poisson level statistics, to chaos, where level statistics become comparable to that of full random matrices.

The band structure of BRMs is observed in the Hamiltonian of various physical systems, such as complex atoms~\cite{Chirikov1985}, isolated thick wire with multiple transverse modes~\cite{Iida1990, Fyodorov1993}, mesoscopic cylinder threaded by magnetic flux~\cite{Gefen1992}, 
and systems of two locally interacting particles confined to a one-dimensional potential well~\cite{Seligman1984, Seligman1985}. Other examples include the one-excitation subspace of many-body systems with long-range couplings~\cite{Richerme2014, Smith2016, Santos2016, Defenu2024, Lerose2023Arxiv, DeTomasi2023,  Buijsman2025arxiv}, field transmission matrix of multimode optical fibers in the weak coupling limit~\cite{Xiong2017}, and models for ocean acoustics~\cite{Hegewisch2012}. The ubiquity of the band structure arises because typical operators are banded when expressed in the ordered eigenbasis of another operator under generic conditions~\cite{Feingold1989, Feingold1991}.
 
BRMs have been applied to the studies of the kicked rotor, where the time evolution operator is banded in the angular momentum representation~\cite{Izrailev1990} with a bandwidth related to the chaos parameter~\cite{Pandey2017, Pandey2020, Kumar2020}. BRMs are also extensively used in transport studies involving systems with local interactions and in the analysis of Anderson localization. This is because BRMs with small bandwidths describe disordered systems with short-range hopping, an example being a chain of one-dimensional harmonic oscillators with random inertia and coupling~\cite{Dyson1953}. In the case of noninteracting Anderson localization~\cite{Herbert1971, Thouless1972}, the systems are described by tridiagonal Hamiltonian matrices, where the off-diagonal elements represent homogeneous nearest-neighbor hopping and the random diagonal elements represent onsite disorder, which leads to spatially localized eigenstates. Also included in the class of BRMs are the tridiagonal matrices of the $\beta$-ensemble~\cite{Dumitriu2002}, where the random off-diagonal elements are distributed according to the $\chi$-distribution. This ensemble presents a nonergodic phase~\cite{Das2024, Das2022b, Das2023, Das2025Arxiv} and nontrivial long-range energy correlations~\cite{Relano2008,Das2022b}.

Despite their widespread applicability, obtaining analytical results for BRMs remains challenging. Unlike full random matrices, the eigenvalues of BRMs are not strongly correlated and their eigenvectors are not fully random vectors. As a result, many studies that involve BRMs rely primarily on numerical calculations.

In this work, we present both numerical and semi-analytical results for BRMs in both the ergodic and nonergodic phases. Using the spectral form factor (SFF), level number variance, and power spectrum of noise, we demonstrate that weak long-range correlations persist in the nonergodic phase, despite the disappearance of short-range correlations.

The SFF enables the analysis of level statistics in the time domain~\cite{MehtaBook,Guhr1998}, quantifying energy correlations across short and long ranges.  It is an effective detector of quantum chaos~\footnote{In this paper, we associate spectral correlations, as observed in full random matrices, with the notion of quantum chaos, although this relationship has been debated.} and has proven to be particularly useful for analyzing molecular spectra~\cite{Leviandier1986, Pique1987, Guhr1990, Hartmann1991, Lombardi1993, Michaille1999}, where line resolution is not as good as in nuclear physics. By obtaining semi-analytical expressions for the SFF of BRMs, we can determine how its timescales depend on the matrix bandwidth. We investigate various characteristic timescales, with an emphasis on the time for the manifestations of spectral correlations (ramp)  and the time for the saturation of the SFF (plateau). 

In the ergodic phase of BRMs, the timescales of the SFF align with those for full random matrices, as expected. But, in the nonergodic phase, the behavior changes significantly, and full random matrices no longer serve as an appropriate reference. In this phase, short-range energy correlations are non-existent and the long-range correlations are fundamentally different from those in full random matrices. We find that, contrary to many-body quantum systems approaching localized phases, where the time for the onset of spectral correlations increases with the disorder strength~\cite{Schiulaz2019, Suntajs2020, Sierant2020}, in BRMs, both timescales, for the appearance of spectral correlations and for the saturation of the SFF, decrease as the bandwidth is reduced. These two timescales eventually merge at the point where the BRM becomes a diagonal matrix and belongs to the Poisson ensemble.

To further investigate the nature of the spectral correlations in the nonergodic phase of BRMs, we analyze the level number variance and the power spectrum of noise, both of which capture short- and long-range energy correlations. Our analysis of the level number variance reveals that in the nonergodic phase, %the Thouless energy, which determines the energy scale below which the energy levels exhibit correlations as in full random matrices, has the same order of magnitude as the mean level spacing, that is, correlations as in full random matrices are nonexistent. Nonetheless, 
weak long-range correlations, distinct from those in full random matrices, persist, which is consistent with  predictions from Ref.~\cite{Erdos2015a}. This finding has implications for the SFF timescales, as mentioned above, and for the power spectrum. We show that the power spectrum can distinguish between the nonergodic and ergodic phases of BRMs at high frequencies of the energy level fluctuations, while at low frequencies, the behavior is identical for both phases.

The paper is organized as follows. The BRM model is introduced in Sec.~\ref{Sec:SFF_BRM}. Our numerical and analytical studies of the evolution of the SFF and its timescales are detailed in Sec.~\ref{Sec:SFF_Tinfty}. Our analyses of the level number variance and power spectrum are presented in Sec.~\ref{Sec:LongRange}.  Concluding remarks are provided in Sec.~\ref{Sec:Conclusion}.

%============================================
\section{Banded Random Matrix Model}
%============================================
\label{Sec:SFF_BRM}

The elements of the $N\times N$ real and symmetric BRMs with bandwidth $b$ that we investigate are defined as
\[
\begin{array}{lll}
H_{ij} \sim {\cal N}(\bar{x},\sigma^2) & \hspace{0.2 cm} \text{for} & \hspace{0.2 cm}  |i - j| \leq b , \\
H_{ij} =0 & \hspace{0.2 cm} \text{for} & \hspace{0.2 cm}  |i - j| >b  ,
\end{array}
\]
where ${\cal N}(\bar{x}, \sigma^2)$ indicates Gaussian random numbers with mean $\bar{x}=0$ and variance $\sigma^2=(1+\delta_{ij})/2$. A BRM reduces to a diagonal matrix, belonging to the Poisson ensemble, when $b = 1$, and becomes a full random matrix from the Gaussian orthogonal ensemble (GOE)~\cite{Dyson1962} when $b=N$.

The typical localization length of the bulk energy eigenstates of BRMs is proportional to $b^2$ provided $1\ll b\ll N$ \cite{Casati1994}. Consequently, the appropriate scaling parameter for studying energy correlations and localization properties is $b^2/N$, so we consider the parametrization
\begin{align}
	b = N^{\frac{\gamma}{2}},\quad \quad 0\leq \gamma \leq 2.
    \label{eq:BRM_param}
\end{align}

In Ref.~\cite{Casati1994}, the localization-delocalization transition was associated with the result for the average localization length $\ell$ given by
\begin{equation}
\ell \propto e^{\langle S \rangle}  \propto
\begin{cases}
         N^{\gamma}, &  \gamma<1 , \\ 
         N, &  \gamma>1 ,
    \end{cases}
    \label{Eq:Entropy}
\end{equation}
where $\langle S \rangle$ is the average of the Shannon information entropy, $S_\alpha=- \sum_{k=1}^N|c_k^\alpha|^2 \ln |c_k^\alpha|^2$, over the eigenstates $|\alpha\rangle$ with coefficients  $c_k^\alpha$ in a chosen basis. The localization length $\propto N$ for $\gamma>1$ is consistent with the ergodic phase of a system hosting energy eigenstates that extend over the entire Hilbert space. On the other hand, for $0<\gamma<1$, the typical localization length goes as $N^\gamma$, which means that the eigenstates are extended but cover only a vanishing fraction of the available Hilbert space, as found in nonergodic phases~\cite{Kravtsov2015, Altshuler2023, Das2022b}. It is only at $\gamma = 0$ that the localization length is $\mathcal{O}(1)$, hence the eigenstates are localized. This is why we find it more appropriate to refer to the transition that happens at the critical point $\gamma_c = 1$ as a ``nonergodic-ergodic transition'' instead of a ``localization-delocalization transition'',  as often found in the literature. We then have the following phases and limiting ensembles for the different values of $\gamma$ in the BRM model: 
%=================================
\begin{table}[h]
\begin{tabular}{|l| l|}
\hline
\hspace{0.1 cm} $\gamma = 0$ \hspace{0.1 cm} & \hspace{0.1 cm} Poisson ensemble, $b = 1$ \hspace{0.1 cm}  \\
\hline
\hspace{0.1 cm} $0<\gamma <1$ \hspace{0.1 cm} & \hspace{0.1 cm}    Nonergodic phase \hspace{0.1 cm} \\
\hline
\hspace{0.1 cm} $\gamma_c = 1$ \hspace{0.1 cm} & \hspace{0.1 cm}    Critical point \hspace{0.1 cm} \\
\hline
\hspace{0.1 cm} $1<\gamma <2$ \hspace{0.1 cm} & \hspace{0.1 cm}    Ergodic phase \hspace{0.1 cm} \\
\hline
\hspace{0.1 cm} $\gamma = 2$ \hspace{0.1 cm} & \hspace{0.1 cm}   GOE, $b=N$ \hspace{0.1 cm} \\
\hline
\end{tabular}
\caption{Limiting ensembles and phases associated with  different values of $\gamma$ in BRM. }
\label{Table:Regimes}
\end{table}
%=================================

The Poisson-Gaudin-Mehta conjecture states that the local bulk statistics of BRMs with $N \rightarrow \infty$ is the same as that of the Poisson ensemble for $\gamma<1$  and of the GOE for $\gamma>1$ \cite{Olver2018}. The transition from one behavior to the other can be verified with the distribution of the spacings $s$ between consecutive unfolded energy levels~\cite{Gomez2002}. When the system size is sufficiently large, $P(s) = e^{-s}$ for the Poisson ensemble ($\gamma=0$), and for the GOE ($\gamma=2$), the Wigner's surmise, $P(s) = \frac{\pi}{2}s e^{-\frac{\pi}{4} s^2}$ is approximately valid.

For a general $b$, $P(s)$ has been studied using supersymmetry~\cite{Spencer2011} and can be well fitted using the Izrailev distribution~\cite{Izrailev1998,Izrailev1990}, 
\begin{align}
	\label{eq_s_Izrailev}
	%P(s,\beta) = As^\beta (1 + B\beta s)^{f(\beta)} e^{-\frac{\pi^2}{16}\beta s^2 - \frac{\pi}{2}\del{1 - \frac{\beta}{2}}s } ,
    P(s,\beta) = A \left( \frac{\pi s}{2}\right)^{\beta} \!\! \exp \left[ -\frac{1}{4} \beta \left( \frac{\pi s}{2}\right)^{2}  - \left( B s -\frac{\beta}{4} \pi s\right) \right]
\end{align}
where $A, B$ are normalizing parameters obtained from
\[
\int_0^{\infty} P(s,\beta) ds =1 \hspace{0.3cm} \text{and} \hspace{0.3 cm} \int_0^{\infty} s P(s,\beta) ds =1 ,
\]
and $\beta$ is the fitting parameter~\footnote{In the case of sparse BRMs, the Brody formula provides a better fit for the level spacing distribution~\cite{Prosen1993}}.
For a random tridiagonal matrix ($b = 2$), the level spacing distribution follows the ansatz $P(s\to 0)\approx s\log^{N-2}(s^{-1})$, which has been verified for small system sizes up to $N = 4$~\cite{Grammaticos1990, Caurier1990}. 

%=================================
\begin{figure}[t]
	\centering
	\includegraphics[width=0.48\textwidth]{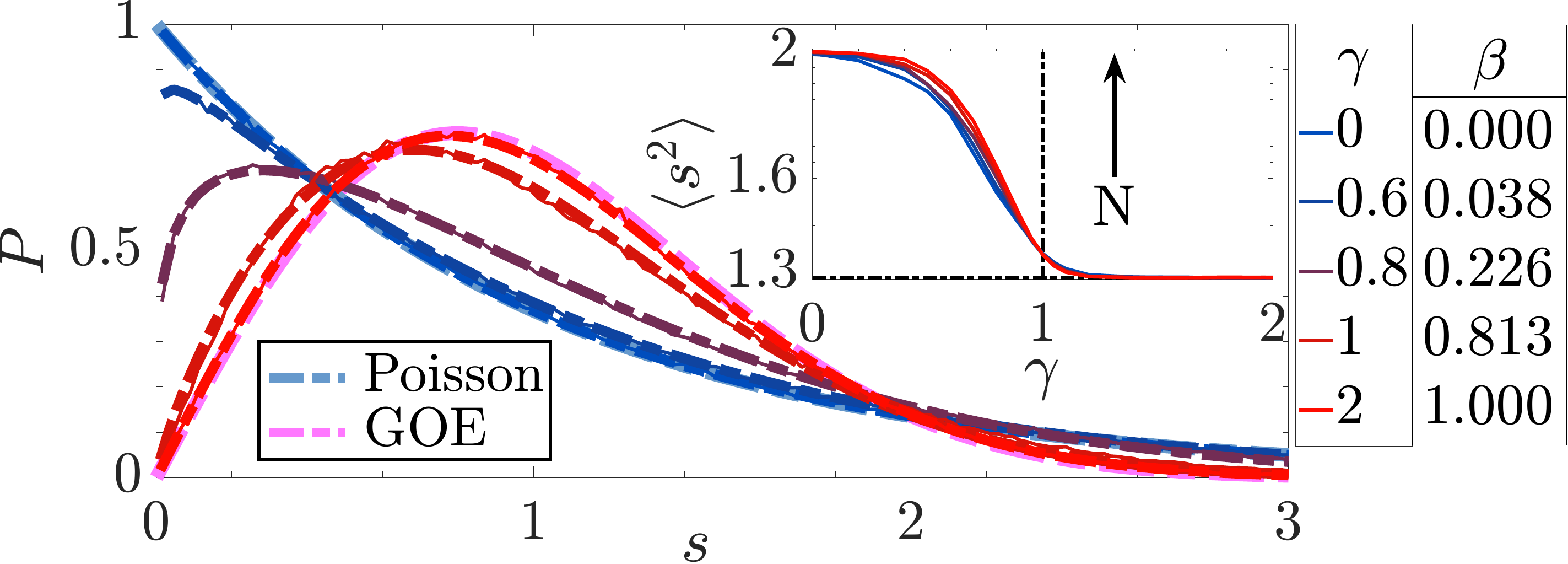}
	\caption{Level spacing distribution for BRMs of various values of $\gamma$, from $\gamma=0$ to $\gamma=2$. Solid lines indicate numerical results and dashed lines represent the fitted curve using the Izrailev distribution~\cite{Izrailev1998} with a fitting parameter $\beta$ provided in the legend. The matrix size is $N = 16384$, average is performed over 128 realizations, and for each realization 40\% of the eigenvalues in the middle of the spectrum are considered. 
		%============
		The inset shows the second moment of the nearest-neighbor level spacing, $\langle s^2\rangle$ vs.~$\gamma$ for matrix sizes from $N = 1024$ to $N=16384$. For the GOE with $N\gg 1$, $\langle s^2\rangle \approx 1.285$~\cite{Das2023a}. 
	}
	\label{fig_NNS}
\end{figure}
%================================

Figure~\ref{fig_NNS} shows that the Izrailev distribution with fitting parameter $\beta$ indeed captures the level spacing distribution for BRMs across all values of $\gamma$. The values of $\beta$ obtained from the resulting fit are system size invariant with respect to $b^2/N$ \cite{Casati1991}. We provide the values of the fit parameter $\beta$ in the legend of Fig.~\ref{fig_NNS}.

In the inset of Fig.~\ref{fig_NNS}, we plot the second moment of the nearest-neighbor spacing $s$ as a function $\gamma$ for different system sizes to show the crossover from  the Poisson ($\langle s^2\rangle = 2$) to the GOE ($\langle s^2\rangle \approx 1.285$ \cite{Das2023a}) limit. At the critical point ($\gamma_c = 1$), the level spacing distribution becomes scale invariant with $\langle s^2\rangle \approx 1.36$, although it does not follow the semi-Poisson statistics~\footnote{A scale invariant spectral statistics different from the semi-Poisson statistics is also observed in the power-law banded random matrices~\cite{Varga2000}.}. For $\gamma<1$, we observe that $\mean{s^2}$ approaches $2$ as $N$ increases, indicating that in the thermodynamic limit ($N\to\infty$), $\gamma_c=1$ marks a transition instead of a crossover and that short-range energy correlations are absent in the nonergodic phase. A detailed analysis of the second and higher moments of $s$ for BRM is provided in~\cite{Cheon1990}. %{\color{blue} As the crossover curves of $\mean{s^2}$ for different values of $N$ intersect at $\gamma = 1$ and get sharper with $N$, $\mean{s^2}_{N\to \infty}\to 2$ for any value of $\gamma<1$. Hence, short-range energy correlation is absent in the nonergodic phase ($\gamma<1$) in the thermodynamic limit ($N\to\infty$).}

An alternative way to detect short-range correlations that avoids the unfolding procedure is the ratio $r$ of consecutive nearest-neighbor spacings~\cite{Atas2013}. The interpolating function to assess the
degree of chaos of a given system was proposed in~\cite{Corps2020} and provides a good fit to the density of $r$ for BRMs with any value of $\gamma$. We have also observed that the mean, $\langle r \rangle $, is system size invariant with respect to $b^2/N$.

%================================
\section{Spectral Form Factor}
\label{Sec:SFF_Tinfty}

%=================================
\begin{figure*}[th]
	\centering
	\includegraphics[width=\textwidth]{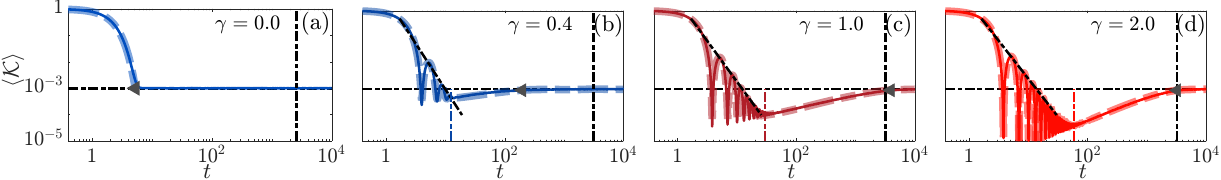}
	\caption{Spectral form factor for BRMs with (a) $\gamma=0$ (Poisson ensemble), (b) $\gamma=0.4$, (c) $\gamma=1$, (d) $\gamma=2$ (GOE),  where $N = 1024$. The solid lines represent numerical results averaged over 1024 realizations and the dashed curves denote the analytical expressions from Eq.~\eqref{eq_SFF_Poisson} for $\gamma = 0$, Eq.~\eqref{eq_SFF_BRM} for $0<\gamma<2$, and Eq.~\eqref{eq_SFF_GOE} for $\gamma = 2$. Horizontal dashed lines give the asymptotic SFF, $\overline{\mathcal{K}} = N^{-1}$. Vertical lines mark the dip ($\tdip$) and Heisenberg ($\tH$) times, where $\tdip<\tH$. The marker $\blacktriangleleft$ indicates the relaxation time, $\tR$. The SFF power-law decay $\propto t^{-3}$ is also indicated with a dashed line in (b)-(d). 
	}
	\label{fig_SFF_Tinfty}
\end{figure*}
%================================ 

The SFF detects both short- and long-range correlations in the energy spectrum, which makes it a useful tool for identifying the transition from the ergodic to the nonergodic phase. It is defined as
\begin{align}
	\label{eq_SFF_infnt}
	\sff{t} = \frac{1}{N^2}\abs{\sum_{n=1}^{N} e^{-iE_n t}}^2 = \frac{1}{N^2}  \sum_{n,m=1}^{N} e^{-i(E_n - E_m) t},
\end{align}
where $\cbr{E_1, E_2,\dots, E_N}$ are the energy levels, $t$ is time, and the normalization factor $N^{-2}$ ensures that $\sff{0}=1$. 

The SFF has been employed in the investigation of scale-invariant critical dynamics~\cite{Hopjan2023, Hopjan2023a} and the stability of the many-body localized phase~\cite{Suntajs2020}.
Its strong connection with the survival probability (probability of detecting the initial state later in time) enables the study of spectral correlations through the dynamics of quantum systems~\cite{Alhassid1992, Campo2011, Campo2016, Torres2017, Torres2017Philo, Torres2018, Campo2017, Campo2018, Schiulaz2019, Xu2021, Lerma2019, Das2024a}. 
The SFF and the survival probability can be experimentally measured~\cite{Joshi2022,Dag2023,Das2024a,Vallejo2025proposalarxiv}, as demonstrated in~\cite{Dong2025}.

The SFF averaged over an ensemble of random matrices can be decomposed in three parts~\cite{MehtaBook, Schiulaz2019},
\begin{align}
	\label{eq_threeParts}
	\mean{\sff{t}} &= \left| \int e^{-iE t} \rho(E) dE \right|^2  & \\
	& - \frac{1}{N(N-1)}  \int e^{-i(E - E') t} T_2(E, E') dE dE' + \sffbar, \nonumber
\end{align}
explained as follows. The first term is the squared modulus of the Fourier transform of the density of states, 
\begin{align}
	\begin{split}
		\rho(E) = \frac{\Tr \delta\del{E - H}}{N} = \frac{1}{N}\sum_{n = 1}^{N}\delta\del{E - E_n},
	\end{split}
\end{align} 
which characterizes the initial decay of the SFF. The second term in Eq.~\eqref{eq_threeParts} contains the two-level cluster function, $T_2(E,E')$ \cite{MehtaBook}, which is only present when the eigenvalues are correlated. The last term, $\sffbar$ in Eq.~\eqref{eq_threeParts}, is the infinite-time average which marks the long-time saturation value of the SFF and $\sffbar = N^{-1}$ for any system.

In Figs.~\ref{fig_SFF_Tinfty}(a)-(d), we show the SFF for BRMs with four different values of $\gamma$, respectively: $\gamma=0$ (Poisson ensemble), $\gamma= 0.4$, $\gamma=1$, and $\gamma=2$ (GOE). The solid lines give numerical results and the dashed lines correspond to the semi-analytical results [Eqs.~\eqref{eq_SFF_BRM}, \eqref{eq_SFF_GOE} and \eqref{eq_SFF_Poisson}, described in Sec.~\ref{Sec:AnalyticalSFF}], indicating excellent agreement.

Using Taylor expansion of the SFF in Eq.~\eqref{eq_SFF_infnt} around $t = 0$, one can see that it has a universal quadratic decay, $\sff{t} \sim 1 -\sigma^2_E t^2$~\cite{Schiulaz2019}, at very short times ($t \ll \tZ=\sigma_E^{-1}$), where 
\begin{align}
	\sigma^2_E = \langle E^2 \rangle - \langle E \rangle^2
\end{align}
is the energy variance of the density of states $\rho(E)$, 
the $n$th energy moment is $\mean{E^n} \equiv \int dE\; E^n\rho(E)$, 
and $\tZ$ is the Zeno time~\cite{Chiu1982}.
Since the spectral width of a BRM depends on $N$, we scale the energy levels as $E_n\to E_n/2\sigma_E$, such that the bulk energy levels are within $\pm 1$ and energy variance becomes 1/4. Such a global scaling is not related to the unfolding of the energy spectrum. Beyond $\tZ$, the behavior of the SFF depends on the shape of $\rho(E)$, as further discussed in Sec.~\ref{Sec:DOS}, and on the spectral correlations, analyzed in Sec.~\ref{Sec:Hole}.  

The effects of the second term in Eq.~\eqref{eq_threeParts} are only relevant in Figs.~\ref{fig_SFF_Tinfty}(b)-(d), because any two eigenvalues in Fig.~\ref{fig_SFF_Tinfty}(a) are uncorrelated. The presence of correlated eigenvalues in Figs.~\ref{fig_SFF_Tinfty}(b)-(d) leads to the dip of $\mean{\sff{t}}$ below $\sffbar$ followed by a ramp toward saturation. This dip-ramp-plateau structure is known as the correlation hole~\cite{Leviandier1986,Pique1987,Guhr1990,Hartmann1991,Alhassid1992,Lombardi1993,Michaille1999,Leyvraz2013,Torres2017,Torres2017Philo,Torres2018,Schiulaz2019,Das2024a,VallejoFabila2024},  which is a definite signature of spectral correlations~\cite{Das2024a}. The correlation hole can be detected experimentally in currently available superconducting quantum processors~\cite{Dong2025} and possibly also in experiments with trapped ions, cold atoms, and existing quantum computers~\cite{Das2024a}.

%%%%%%%%%%%%%%%%%%
\subsection{Spectral Form Factor Analytical Expression}
%%%%%%%%%%%%%%%%%%
\label{Sec:AnalyticalSFF}

To derive the analytical results for the SFF shown in Fig.~\ref{fig_SFF_Tinfty}, one needs the analytical expressions for the density of states and for the correlation hole of BRMs with different values of $\gamma$. %

%=================================================
\subsubsection{Density of states}
\label{Sec:DOS}

The density of states can be obtained in terms of the energy moments. In an ensemble of BRM, the probability density in the matrix space is normalized while the matrix norm, $\sqrt{\Tr H^2}$, and the norm of the off-diagonal part have finite ensemble averages. Using the maximum entropy principle~\cite{Balian1968, Jaynes1957, Das2022, Das2022a}, i.e.~by maximizing the Shannon entropy in the matrix space subject to the above constraints, we get the density of BRM in the matrix space as
\begin{align}
	\label{eq_H_density}
	P(H) = \frac{1}{\mathcal{Z}} \exp\del{-\frac{1}{2}\sum_{j = 1}^{N} H_{ii}^2 - \sum_{0 < i-j < b}^{N} H_{ij}^2 }
\end{align}
where $\mathcal{Z} = 2^{\frac{N}{2}} \pi^{\frac{b(2N+1-b)}{4}}$ is the normalization constant. Equation~\eqref{eq_H_density} enables the computation of the energy moments, $\mean{E^n} = \frac{1}{N} \int dH P(H) \Tr H^{n}$. 

With the energy moments, we can express the density of states as
\begin{align}
	\rho(E) = \frac{1}{\sqrt{2\pi}} \sum_{n = 0}^{\infty} \frac{\mean{E^{2n}}}{2n!}  \dfrac{d^{2n} \delta(E)}{dE^{2n}} .
\end{align}
Even though the  odd energy moments are zero for BRM, reflecting the symmetry of the density of states about $E = 0$, the explicit computation of all even energy moments from $P(H)$ is a daunting task. However, the density of states of the Rosenzweig-Porter ensemble (RPE)~\cite{Rosenzweig1960, Kravtsov2015, Facoetti2016, Monthus2017, Bogomolny2018, Das2019, Soosten2019, Buijsman2022, Venturelli2023, DeTomasi2022, Das2023a, Sarkar2023}, has been obtained earlier~\cite{Bertuola2005}. We assume that $\rho(E; \alpha, a)$ for RPE approximates the density of states for the BRM and derive (see details in App.~\ref{sec_BRM_moms})
\begin{align}
	\label{eq_DOS_BRM}
	\begin{split}
		\rho(E, \kappa) &= \frac{ \sqrt{\frac{8(\kappa+1)}{\pi^3}} }{e^{2(\kappa+1)E^2}} \int_{-1}^{1} dx \frac{\sqrt{1-x^2}}{ e^{2\kappa x^2 + 4\sqrt{\kappa^2+\kappa} Ex} },
	\end{split}
\end{align}
where

\begin{align}
	\label{eq_kappa}
    \begin{split}
		\kappa &\equiv \frac{1 - K + \sqrt{1-K}}{K}
	\end{split}
\end{align}
and 
\begin{align}
	\label{eqA_dos_kurt}
	K = \frac{\mean{E^4}}{\mean{E^2}^2} - 2.
\end{align}
is the shifted Kurtosis, which is 1 for a Gaussian distribution and 0 for the semi-circle. 

For $\kappa = 0$, Eq.~\eqref{eq_DOS_BRM} yields the Gaussian distribution $\sqrt{\frac{2}{\pi}} e^{-2E^2}$, valid for the Poisson ensemble ($\gamma = 0$). For $\kappa\to \infty$,  we get Wigner's semi-circle law $\rho_{\mathrm{GOE}}(E) = \frac{2}{\pi} \sqrt{1 - E^2}$ \cite{Backer1995, Casati1993a}, valid for the GOE ($\gamma = 2$).

%=================================
%	DOS
\begin{figure}[t]
	%\centering
	\includegraphics[width=0.45\textwidth]{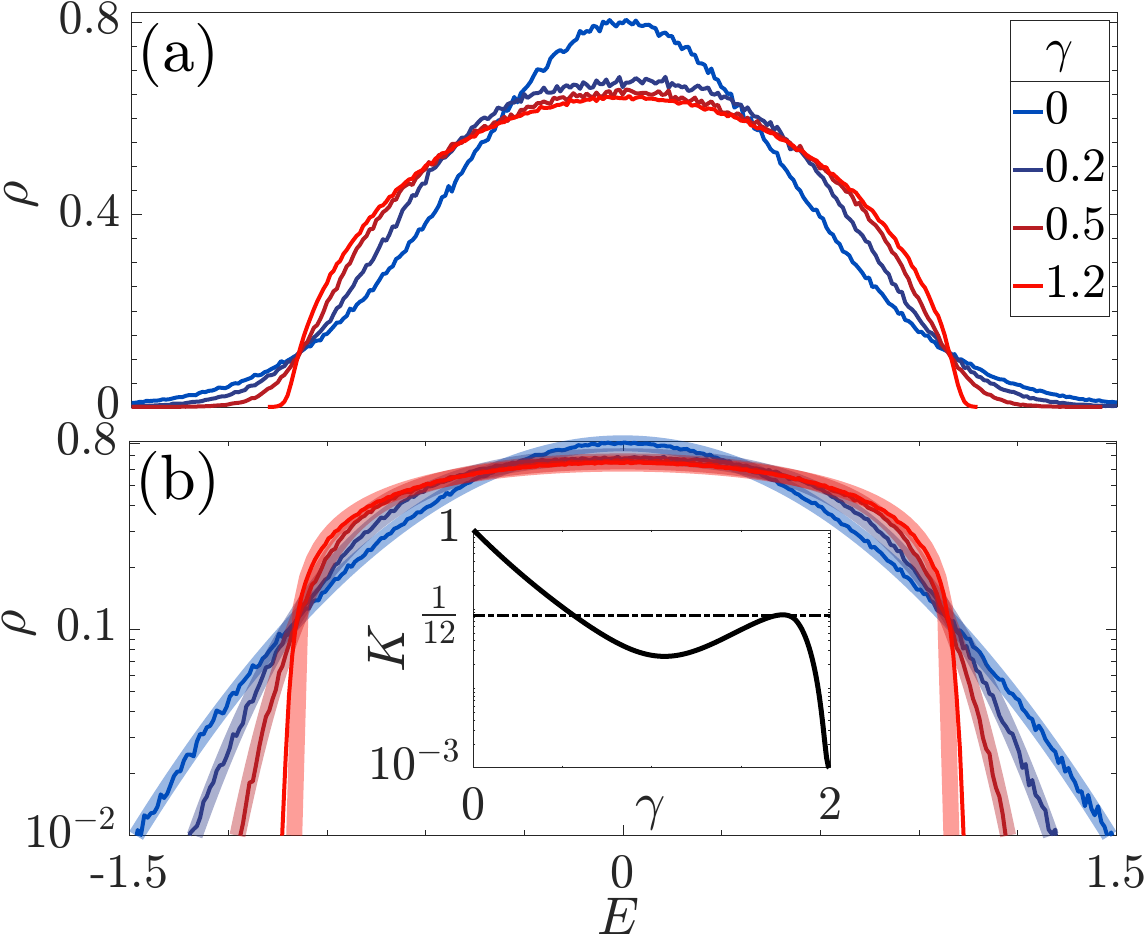}
	\caption{Density of states for BRMs with various values of $\gamma$ in (a)~linear scale and (b)~log-scale in the y-axis; $N = 1024$ and average over 2048 realizations. Numerical results are shown with solid lines and the shaded curves in panel (b) denote $\rho(E, \kappa)$ from Eq.~\eqref{eq_DOS_BRM}. The shape of $\rho(E)$ is Gaussian for $\gamma = 0$ (Poisson ensemble) and semicircular for $\gamma = 2$ (GOE). The GOE curve is not shown, because the result for $\gamma=1.2$ is already very close to it. The inset of panel (b) shows the shifted kurtosis vs.~$\gamma$ for system size $N = 1024$ [see Eq.~\eqref{eqA_dos_kurt}].
	}
	\label{fig_DOS}
\end{figure}
%=================================

Figure~\ref{fig_DOS} compares numerical results  (solid lines) for the density of states of BRMs of various values of $\gamma$ with  
Eq.~\eqref{eq_DOS_BRM} (shaded curves), confirming that the equation provides an excellent approximation for $\rho(E)$. The density of states in Fig.~\ref{fig_DOS} is centered at $\mean{E} = 0$ and has a standard deviation equal to $1/2$ irrespective of $\gamma$ due to the global scaling $E_n\to E_n/2\sigma_E$. For the Poisson ensemble, the density of states in Fig.~\ref{fig_DOS} is a Gaussian function and for the GOE, we recover the semicircle law.

\iffalse
In the thermodynamic limit ($N\to\infty$), $\rho_\text{GOE}(E)$ holds for any $\gamma>0$. For finite $N$ and $b\propto N$, the semicircular shape $\rho_\text{GOE}(E)$ is still approximately valid~\cite{Casati1993a, Molchanov1992}, apart from deviations~\cite{Guionnet2002, Li2013,Jana2016, Olver2018} that become maximum around $b = 2N/5$~\cite{Olver2018}. Indeed, the plot for the kurtosis as a function of $\gamma$ in the inset of Fig.~\ref{fig_DOS} shows that $K$ exhibits a local maximum for the value of $\gamma$ leading to $b \sim 2N/5$. At this point, $K_{N\to \infty}=1/12$~(see App.~\ref{sec_BRM_DOS}).

For finite $N$ and non-extensive $b$ in $N$, the BRM density of states has a shape between Gaussian and semicircle depending on $\gamma$, as seen in Fig.~\ref{fig_DOS} for $\gamma<1$. This is also observed in the inset of Fig.~\ref{fig_DOS}, where $K \rightarrow 1$ (Gaussian shape) as $\gamma$ decreases. The inset also presents a local minimum at $\gamma=1$. At this point, $K=0$ for $N\rightarrow \infty$ (see App.~\ref{sec_BRM_DOS}).
\fi

%=================================
\subsubsection{Correlation hole}
\label{Sec:Hole}

The two-level cluster function, $T_2(E, E')$, captures the correlations among the energy levels~\cite{MehtaBook, Bogomolny2001}. For unfolded energy levels, $\cbr{\Ecal_j}$, the two-level cluster function is denoted by $Y_2(\Ecal, \Ecal')$. Given the mean level spacing $\mu$, the above functions are related as $T_2(E, E') \approx Y_2(\Ecal, \Ecal')/\mu^2$. Since $Y_2(\Ecal, \Ecal') = Y_2(\Delta_{\Ecal})$ \cite{MehtaBook, Buijsman2020}, where $\Delta_\Ecal = |\Ecal - \Ecal'|$ is the spacing of two unfolded energy levels, we can express the Fourier transform of the two-level cluster function in Eq.~(\ref{eq_threeParts}) as
\begin{align}
\label{eq_def_b2}
	\begin{split}
		&-\frac{1}{N(N-1)} \int dE \int dE' e^{-i2\pi(E-E')t} T_2(E, E')\\
		&\approx -\frac{1}{N(N-1)} \int \mu d\Ecal \int \mu d\Ecal' e^{-i2\pi\mu (\Ecal - \Ecal')t} \frac{1}{\mu^2} Y_2(\Ecal, \Ecal')\\
		&\approx -\frac{1}{N} \int d\Delta_\Ecal\: e^{-i2\pi\mu \Delta_\Ecal t} Y_2(\Delta_\Ecal) \\
		&= -\frac{1}{N} b_2(\mu t),
	\end{split}
\end{align} 
where $b_2$ is the two-level form factor. This function controls the long-time behavior of the SFF~\cite{Schiulaz2019}, giving rise to the correlation hole if the eigenvalues are correlated. 

For the Poisson ensemble, $b_2(\tau) = 0$, where $\tau$ is the dimensionless time. For GOE~\cite{MehtaBook}, 
\begin{align}
	\label{eq_b2_GOE}
	b_2^{\mathrm{GOE}}(\tau) &= \begin{cases}
		1 - 2\tau + \tau\log\del{1 + 2\tau}, & \tau\leq 1,\\
		\tau\log \left( \dfrac{2\tau + 1}{2\tau - 1} \right) - 1, &\tau > 1 .
	\end{cases}
\end{align}
Motivated by the two-level form factor of the Rosenzweig-Porter ensemble~\cite{Pandey1995}, we propose the following ansatz for BRM, 
\begin{align}
	\label{eq_b2_fit}
	b_2(\tau) = b_2^{\mathrm{GOE}}(\tau) + f(\tau) e^{-\frac{\tau}{\eta}} , 
\end{align}
where $f(\tau)$ is a second order polynomial and $\eta$ is a fitting parameter.

\begin{figure}[t]
	\centering
	\includegraphics[width=0.9\columnwidth]{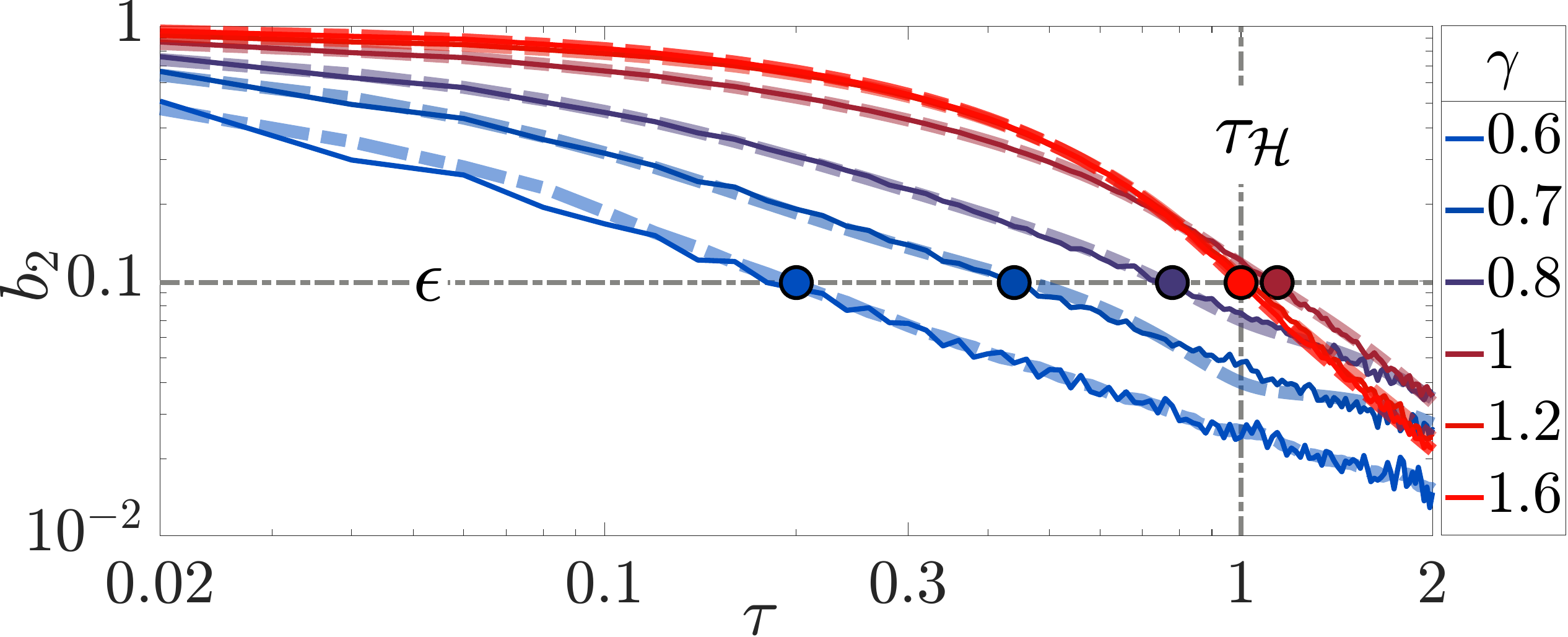}
	\caption{Two-level form factor as a function of the dimensionless time $\tau$ for BRMs at various values of $\gamma$; $N = 1024$ and average over 2048 realizations. Solid lines show numerical results and dashed lines represent the ansatz in Eq.~\eqref{eq_b2_fit} with fitting parameters given in Table~\ref{tab:fit_param_b2}. The vertical dashed line marks the dimensionless Heisenberg time $\tau_{\mathrm{H}}=1$ and the horizontal dashed line is the tolerance value $\epsilon$ used  to obtain the dimensionless relaxation time $\tau_{\mathrm{R}}$. The circles indicate $\tau_{\mathrm{R}}$ for various values of $\gamma$ while $\tau_{\mathrm{R}} = \tau_{\mathrm{H}} = 1$ for $1<\gamma\leq 2$.
	}
	\label{fig_BRM_two-pt}
\end{figure}

In Fig.~\ref{fig_BRM_two-pt}, we show that the ansatz in Eq.~\eqref{eq_b2_fit} (dashed lines) agrees very well with the numerical results (solid lines). In Table~\ref{tab:fit_param_b2}, we show the fitting parameters used in Fig.~\ref{fig_BRM_two-pt} for various values of $\gamma$. 
Notice that in the ergodic phase ($\gamma>1$), $b_2(\tau) \approx b_2^{\mathrm{GOE}}(\tau)$, with small deviations appearing for $\tau\ll 1$. On the other hand, the second term in Eq.~\eqref{eq_b2_fit} is dominant in the nonergodic phase ($\gamma<1$).
 
%=================================
%	2-point correlation
%=================================

%=================================
\begin{table}[h]
	\begin{tabular}{|c|c|c|c|c|}
		\hline
		$\gamma$ & $\eta$ & a      & b       & c      \\ \hline
		0.6      & 0.2357   & 2.4552 & 0.0799  & 3.0127 \\ \hline
		0.7      & 0.2298   & 7.2859 & 0.0255  & 1.6055 \\ \hline
		0.8      & 0.2711   & 4.3480 & 0.0361  & 1.2508 \\ \hline
		1        & 0.3027   & 2.0196 & 0.0672  & 0.7757 \\ \hline
		1.2      & 0.3682   & 0.0145 & 9.1863  & 0.5486 \\ \hline
		1.6      & 0.0188   & 0.0673 & 13.0405 & 0.0774 \\ \hline
	\end{tabular}	\caption{Fitting parameters for the two-level form factor ansatz in Eq.~\eqref{eq_b2_fit} for different values of $\gamma$ shown in Fig.~\ref{fig_BRM_two-pt}. The 2nd order polynomial has the form $f(\tau) = a(\tau+b)(c-\tau)$.}
	\label{tab:fit_param_b2}
\end{table}
%=================================

%=================================
\subsubsection{Analytical expression: SFF}

Combining the results for the density of states in Sec.~\ref{Sec:DOS} and for the two-level form factor in Sec.~\ref{Sec:Hole}, we evaluate Eq.~\eqref{eq_threeParts} and obtain the following semi-analytical expression for the SFF of BRM, 
\begin{align}
	\label{eq_SFF_BRM}
	\begin{split}
		\mean{\sff{t, \kappa}} &\approx \frac{\fbslj{1}{ \frac{2\sigma_E t}{\sqrt{1+\frac{1}{\kappa}}} }^2}{\frac{\sigma_E^2 t^2}{1+\frac{1}{\kappa}} e^{\frac{\sigma_E^2 t^2}{1+\kappa}}} - \frac{1}{N} b_2\del{ \frac{t}{\tH} }  + \frac{1}{N} ,
	\end{split}
\end{align}
where the first term comes from the Fourier transform of Eq.~\eqref{eq_DOS_BRM} and contains $\fbslj{1}{t}$, which is the Bessel function of the first kind of order 1. The second term is the two-level form factor proposed in Eq.~\eqref{eq_b2_fit}, where the dimensionless time $\tau = \frac{t}{\tH}$ involves the Heisenberg time, $\tH$, given by the inverse of the mean level spaicng, $1/\mu$, and explained below in Eq.~\eqref{eq:tH}. 

For $\gamma > 1$, since $\kappa\gg 1$, Eq.~\eqref{eq_SFF_BRM} agrees well with the SFF of the GOE~\cite{Schiulaz2019, Das2024a}
\begin{align}
	\label{eq_SFF_GOE}
	\mean{\mathcal{K}^{\mathrm{GOE}}(t)} = \frac{\fbslj{1}{2\sigma_E t}^2}{\sigma_E^2 t^2} - \frac{1}{N} b_2^{\mathrm{GOE}}\del{ \frac{t}{ \tH } } + \frac{1}{N},
\end{align}
where $b_2^{\mathrm{GOE}}(\tau)$ is the two-level form factor of the GOE in Eq.~\eqref{eq_b2_GOE}.

For the Poisson ensemble ($\gamma=0$), where the correlation hole does not exist, the analytical expression for the SFF is simply~\cite{Torres2014b, Torres2014c, Tavora2016}
\begin{align}
	\label{eq_SFF_Poisson}
	\begin{split}
		\mean{\mathcal{K}^{\mathrm{P}}(t)} &= \Theta\del{\tR^{\textrm{P}} - t} e^{-\sigma_E^2 t^2} + \Theta\del{t - \tR^{\textrm{P}}} \frac{1}{N} ,
	\end{split}
\end{align}
where $\Theta(x)$ is the Heaviside step function and 
\begin{equation}
	\tR^{\textrm{P}}= \sqrt{\ln N}/\sigma_E
	\label{eq:tR_P}
\end{equation}
is the time when the Gaussian decaying curve of the SFF meets the saturation value $\sffbar=1/N$. We get Eq.~\eqref{eq_SFF_Poisson} from Eq.~\eqref{eq_SFF_BRM} for $\kappa = 0$, since $\lim\limits_{x\to 0} \frac{\fbslj{1}{x}}{x} = \frac{1}{2}$.

%===========================
\subsection{Timescales}

The shortest timescale of the system is the Zeno time $\tZ=\sigma_E^{-1}$, as mentioned in the beginning of Sec.~\ref{Sec:SFF_Tinfty}.
The largest timescale is the Heisenberg time, defined as the inverse of the mean level spacing around $E = 0$,
\begin{equation}
    \tH = \frac{N}{\sigma_E \rho(0, \kappa)} = \frac{\sqrt{\pi} N e^\kappa}{\sigma_E \sqrt{2(\kappa+1)} \del{\fbsli{0}{\kappa} + \fbsli{1}{\kappa}}} ,
    \label{eq:tH}
\end{equation}
where $\kappa$ is given in Eq.~\eqref{eqA_DOS_BRM_1}. 
Note that the global mean level spacing is $\sigma_E/ N$, but the density of states is not uniform for BRMs, with the largest density occurring around $E = 0$ and denoted by $\rho(0,\kappa)$. The energy spread in the vicinity of $E=0$ is thus approximately $\sigma_E \rho(0,\kappa)$, so the local mean level spacing is $\sigma_E \rho(0,\kappa)/N$ and its inverse leads to the Heisenberg time in the equation above.
For the Poisson ensemble ($\gamma=0$), where $\kappa=0$, Eq.~(\ref{eq:tH}) leads to $\tH^{\textrm{P}} = \sqrt{\pi} N/\sqrt{2}\sigma_E$. For the GOE ($\gamma=2$), where $\kappa \rightarrow \infty$, Eq.~(\ref{eq:tH}) gives $\tH^\mathrm{GOE} = \pi N/2\sigma_E$.

Before deriving the intermediate timescales $t\in[\tZ,\tH]$ for BRMs, we discuss them for the Poisson ensemble and the GOE.

The simple behavior of the spectral form factor for the Poisson ensemble in Eq.~(\ref{eq_SFF_Poisson}) implies that beyond the Zeno time, the only remaining timescale is that for the relaxation of $\mean{\mathcal{K}^{\mathrm{P}}(t)}$, which happens at $\tR^{\textrm{P}}$ [Eq.~(\ref{eq:tR_P})]. This time is shorter than the Heisenberg time, $\tR^{\textrm{P}} \ll \tH^{\textrm{P}}$.

For the GOE (and any ensemble showing a correlation hole), beyond the Zeno time and before relaxation, there is the time for the beginning of the ramp, denoted by $\tdip$. This time happens after the power-law decay of the SFF, which is $\propto t^{-3}$ for GOE [see Fig.~\ref{fig_SFF_Tinfty}(d)]  and characterizes the asymptotic behavior of $\del{\fbslj{1}{t}/t}^2$. This algebraic decay is caused by the bounds in the density of states~(\cite{Khalfin1958, Tavora2016, Tavora2017} and the references therein). To obtain $\tdip$, one needs to expand the first term in Eq.~(\ref{eq_SFF_GOE}) for long times $\del{\sim \frac{1}{\pi \sigma_E^3 t^3}}$ and the second term for short times $\del{\sim \frac{1}{N}\del{\frac{4\sigma_E t}{\pi N}-1}}$ and take their derivatives to determine where the two functions meet~\cite{Schiulaz2019}, which happens at
\begin{align}
	\label{eq_tdip_GOE}
	\begin{split}
		\tdip^{\mathrm{GOE}} \approx 3^{\frac{1}{4}}\sqrt{N}/ \sqrt{2}\sigma_E.
	\end{split}
\end{align}
At this time~\cite{Das2024a}, 
$$\mean{\mathcal{K}^{\mathrm{GOE}}\del{t = \tdip^{\mathrm{GOE}}} } \approx \frac{2\sqrt{2}}{\pi \sqrt{N^3} } \del{ 3^{\frac{1}{4}} + 3^{-\frac{1}{4}} } .$$

For $t>\tdip^{\mathrm{GOE}}$, the SFF is dictated by the two-level form factor up to the saturation at $\tR^{\mathrm{GOE}}$. This time can be obtained by expanding $b_2^{\mathrm{GOE}}(t)$ for long times~\cite{Schiulaz2019}, which gives 
\begin{align}
	\label{eq_tR_GOE}
	\begin{split}
		\tR^{\mathrm{GOE}} = \frac{\pi N}{4\sigma_E \sqrt{3 \epsilon}} .
	\end{split}
\end{align}
In the equation above, $\epsilon$ is a small tolerance value used because the SFF approaches equilibrium in a power-law manner,  $b_2^{\mathrm{GOE}}(t \rightarrow \infty) \propto t^{-2}$. Since $\tH^{\mathrm{GOE}} = \pi N/2\sigma_E$, Eq.~(\ref{eq_tR_GOE}) implies that the SFF relaxes on the same timescale as the Heisenberg time, $\tR^{\mathrm{GOE}}\approx \tH^{\mathrm{GOE}} $. This is confirmed in Fig.~\ref{fig_SFF_Tinfty}(d), 
where the vertical line indicating the Heisenberg time coincides with the marker $\blacktriangleleft$  representing the relaxation time.

%=====================================

Similar to the GOE, the SFF of BRMs with $0<\gamma<2$ exhibits a ramp beginning at $\tdip$ and eventually saturates at $\tR$. These two timescales are discussed next. A key finding of our work is the existence of a correlation hole in the nonergodic phase despite the absence of short-range correlation with a $\tdip$ that decreases as $\gamma$ decreases, as further elaborated below.

\subsubsection{Time for the beginning of the ramp for BRM: $\tdip$}

To obtain $\tdip$ for BRMs with $0<\gamma <2$, we expand the first term in Eq.~\eqref{eq_SFF_BRM} for large times and $b_2(\tau)$ for short times, as done in the derivation of Eq.~(\ref{eq_tdip_GOE}). 

The expansion of the first term  in Eq.~\eqref{eq_SFF_BRM} for  $t\gg \frac{\sqrt{1+\kappa^{-1}}}{\sigma_E}$ gives
\begin{equation}
    \frac{\left(1 + \kappa^{-1} \right)^{\frac{3}{2}} e^{-\frac{\sigma_E^2 t^2}{1+\kappa}} \cos
   ^2\left( \frac{\pi }{4} +  \frac{2\sigma_E t}{\sqrt{ 1 + \kappa^{-1} }}\right)}{\pi  \sigma_E^3 t^3} ,
\end{equation}
which implies oscillations with the envelope $t^{-3} e^{-\frac{\sigma_E^2 t^2}{1+\kappa}}$. As one approaches the GOE and $\kappa\gg 1$, the Gaussian part of this expression becomes negligible, since we are considering finite times up to $t\lesssim \tdip$. In this case, the SFF exhibits a power-law decay $\propto t^{-3}$. In fact, we verified numerically that even in the nonergodic region with $0<\gamma<1$,  the power-law decay exponent is still close to 3, as seen in Fig.~\ref{fig_SFF_Tinfty}(b). This means that variations in the value of $\tdip$ with respect to $\tdip^\text{GOE}$ are mainly caused by deviations of the BRM two-level form factor $b_2(\tau)$ from $b_2^\text{GOE}(\tau)$.

The differences between $b_2(\tau)$ and $b_2^\text{GOE}(\tau)$ for $\tau \ll 1$ are evident in Fig.~\ref{fig_BRM_two-pt} for any $\gamma<2$. This deviation is particularly pronounced in the nonergodic phase and is attributed to the presence of weak long-range correlations (further discussed in the next section). It results in the peculiar situation where the correlation hole persists in the nonergodic phase despite the absence of short-range energy correlations. In this phase, we observe that the depth of the correlation hole diminishes as $\gamma$ decreases, resulting in shorter values of $\tdip$ [compare Fig.~\ref{fig_SFF_Tinfty}(d) with Figs.~\ref{fig_SFF_Tinfty}(b)-(c)]. 

The behavior of $\tdip$ for BRMs is markedly different from what is observed in disordered many-body quantum systems, which deserves further explanations. In many-body quantum systems, disorder can take the system away from the chaotic phase. By increasing the disorder strength, the length scale of energy correlations (Thouless energy) is reduced~\cite{Garcia2016}, delaying the onset of the ramp~\cite{Schiulaz2019,Suntajs2020}. However, any two eigenvalues with spacing smaller than the Thouless energy remain correlated as in full random matrices. This behavior is similar to what  is seen in the ergodic phase ($\gamma>1$) of BRM, where $\tdip$ is close to $\tdip^{\mathrm{GOE}}$ and the ramp of the SFF is mainly described by the two-level form factor of GOE. But everything changes in the nonergodic phase ($\gamma<1$) of BRM, where 
the Thouless energy is comparable to the mean level spacing. This means that the existing weak long-range correlations causing the correlation hole in Fig.~\ref{fig_SFF_Tinfty}(b) are fundamentally different from that of the GOE. As $\gamma$ decreases, the depth of the correlation hole diminishes, yet the power-law decay in the slope of the SFF remains $\propto t^{-3}$, leading to the reduction of $\tdip$.

%=================================
%	Different timescales for SFF
\begin{figure}[t]
	\centering
	\includegraphics[width=0.45\textwidth]{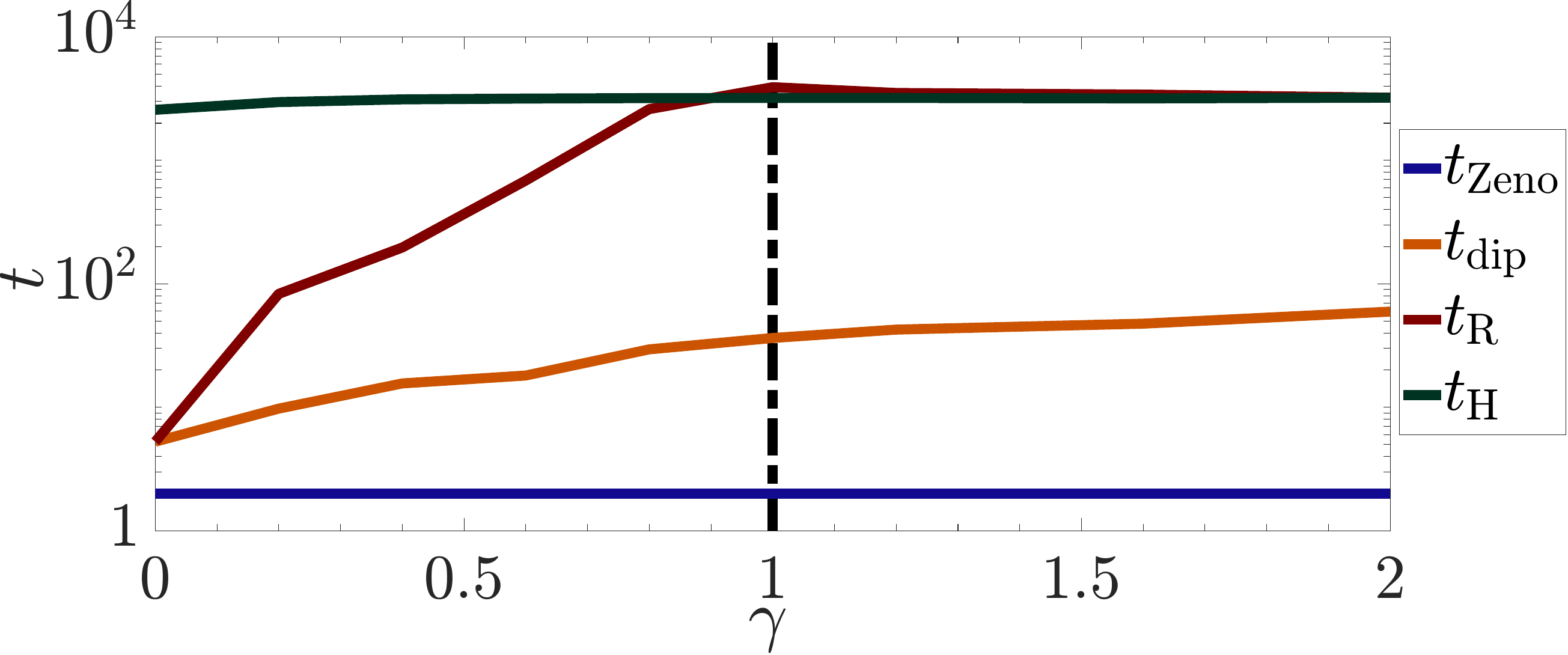}
	\caption{Zeno time ($\tZ$), time for the beginning of the ramp ($\tdip$), relaxation time ($\tR$), and Heisenberg time ($\tH$) as a function of $\gamma$; $N = 1024$. The spectrum is scaled such that $\sigma_E = 1/2$ for all values of $\gamma$.
	}
	\label{fig_BRM_t_SFF_1024}
\end{figure}
%=================================

%=================================
\subsubsection{Relaxation time for BRM: $\tR$}

To obtain $\tR$ for BRMs with $0<\gamma <2$, we expand $b_2(\tau)$ for long times, as done in the derivation of Eq.~(\ref{eq_tR_GOE}).
For $\gamma>1$,  $b_2(\tau)$ in Eq.~\eqref{eq_b2_fit} is similar to $b_2^{\text{GOE}} (\tau)$ for large $\tau$, as visible in Fig.~\ref{fig_BRM_two-pt}, 
and $\tR\approx \tH$.

As $\gamma$ decreases below 1 and the second term in Eq.~\eqref{eq_b2_fit} becomes dominant, the two-level form factor gets exponentially suppressed, as seen in Fig.~\ref{fig_BRM_two-pt}. In this case, the correlation hole gets shallower and $\tR$ becomes smaller than $\tH \sim \mathcal{O}(N)$, gradually approaching $\tR^{\mathrm{P}}$ given in Eq.~\eqref{eq:tR_P}. This change can be observed by comparing Fig.~\ref{fig_SFF_Tinfty}(d) for $\gamma=2$ (GOE) with  Fig.~\ref{fig_SFF_Tinfty}(b) for $\gamma=0.4$. In the latter case, the marker $\blacktriangleleft$,  representing the relaxation time, is seen before the Heisenberg time.

Figure~\ref{fig_BRM_t_SFF_1024} summarizes our discussions about the timescales involved in the evolution of the SFF. The Zeno time ($\tZ$), the time for the beginning of the ramp ($\tdip$), the relaxation time ($\tR$), and the Heisenberg time ($\tH$) are shown as a function of $\gamma$ for $0\leq \gamma \leq2$. The most evident changes happen in the nonergodic phase, where $\tdip$ and $\tR$ decrease as $\gamma$ approaches zero, eventually merging together when the correlation hole vanishes.

Thus, we find that the SFF of a BRM in the nonergodic phase ($0 < \gamma < 1$) exhibits a correlation hole (Figs.~\ref{fig_SFF_Tinfty}(b)-(d)) despite the absence of short-range energy correlations. This is a manifestation of the weak long-range correlations following Altshuler-Shklovskii statistics, as we discuss next.

%============================================
\section{Long-range spectral statistics of BRM}\label{Sec:LongRange}
%============================================

In this section, we investigate two measures that capture short- as well as long-range energy correlations, the level number variance and the power spectrum, and we identify the Thouless energy, $\Eth$~\cite{Thouless1974, Altshuler1986, Altshuler1988, Bertrand2016, Serbyn2017, Sutradhar2019, Sierant2019}. In the ergodic phase, the Thouless energy determines the energy scale  %{\color{blue} separating the short- and long-range energy correlation behaviors. In particular, in the ergodic phase, below the Thouless energy scale, any two energy levels are strongly correlated as in GOE.} 
below which any two energy levels are correlated as in the GOE.

%============================================
\subsection{Level Number Variance}
%============================================

The level number variance is a tool to study energy correlations at length $\Delta_\Ecal$, providing information about the rigidity of the spectrum. It is
defined as~\cite{Guhr1998},
\begin{equation}
	\label{eq_nvar}
	\Sigma^2(\Delta_\Ecal) = \langle \mathcal{N}^2(\Delta_\Ecal, \Ecal) \rangle - \langle  \mathcal{N}(\Delta_\Ecal, \Ecal)\rangle^2 ,
\end{equation}
where $\Ecal$ is the unfolded energy, $\mathcal{N}(\Delta_\Ecal, \Ecal)$ is the number of energy levels in the window $[\Ecal - \Delta_\Ecal/2, \Ecal + \Delta_\Ecal/2]$, and $\mean{.}$ indicates average over the spectrum. 

For GOE ($\gamma=2$), the energy spectrum is rigid, so any two energy levels are correlated and the level number variance exhibits a logarithmic behavior with $\Delta_\Ecal$~\cite{MehtaBook}. 
For the Poisson ensemble ($\gamma=0$), absence of energy correlation leads to the linear increase of $\Sigma^2$ with $\Delta_\Ecal$.

\begin{figure}[t]
	\centering
	\includegraphics[width=0.85\columnwidth]{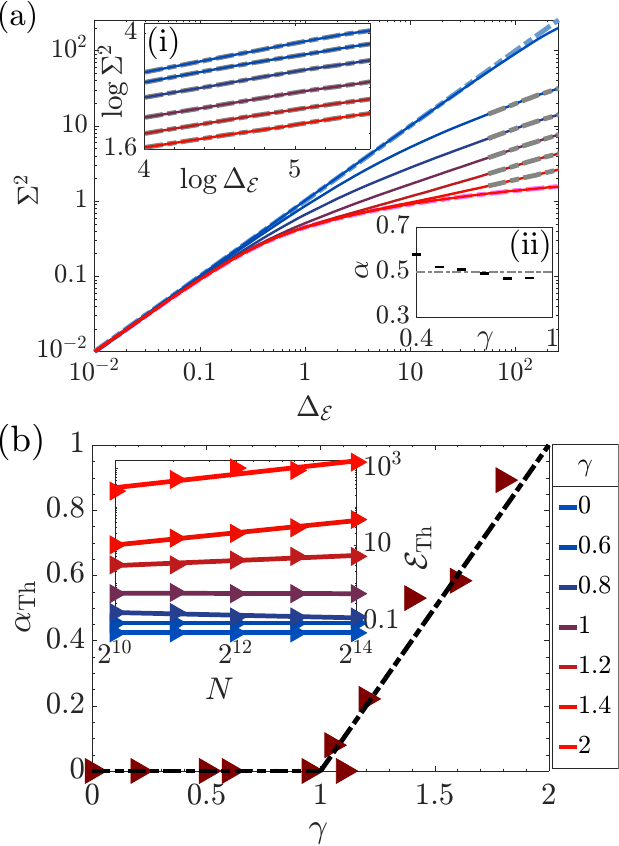}
	\caption{(a) Level number variance for various values of $\gamma$; $N = 1024$, average over middle of the spectrum and ensemble realizations. Dashed blue (pink) line denotes the analytical expression for the Poisson ensemble (GOE). The power-law behavior for large energy gaps, $\Sigma^2(\Delta_\Ecal\gg 1) \propto \Delta_\Ecal^{\alpha}$, is shown with gray dashed lines in the main panel and plotted in the inset (i) along with a linear fit in log-log scale. The fitting is used to obtain the power-law growth exponent, $\alpha$, shown in the inset (ii) as a function of $\gamma$ via errorbars with 95\% confidence interval.
		%==========================
		(b) $\alpha_{\mathrm{Th}}$ from the system size scaling of the unfolded Thouless energy, $\Ecal_{\mathrm{Th}}\propto N^{\alpha_{\mathrm{Th}}}$, as a function of $\gamma$. The dashed line follows Eq.~\eqref{eq_uE_Th}. The inset shows $\Ecal_{\mathrm{Th}}$ vs.~$N$ for various $\gamma$ with linear fit in log-log scale.
	}
	\label{fig_num_var}
\end{figure}

To study the level number variance of BRMs with $0<\gamma<2$, we focus on the middle of the unfolded spectrum, $\Ecal \sim N/2$, and perform ensemble averages. The results in Fig.~\ref{fig_num_var}(a) show curves for $\Sigma^2(\Delta_\Ecal)$ between those for the Poisson ensemble and the GOE.

In the ergodic phase of BRM ($\gamma>1$), an excitation propagates diffusively with a diffusion constant $\sim b^2$~\cite{Fyodorov1991}, hence, $\Eth \approx b^2/N^2$ provided $\sigma_E$ is system size independent~\cite{Erdos2015a}. In contrast, in the non-ergodic phase, the absence of any short-range energy correlation implies that the Thouless energy has the same order of magnitude as the mean level spacing, thus $\Eth \sim \sigma_E/N$. This means that the unfolded Thouless energy can be expressed as
\begin{align}
	\label{eq_uE_Th}
	\Ecal_{\mathrm{Th}} = N^{\alpha_{\mathrm{Th}}},\quad \alpha_{\mathrm{Th}} = \begin{cases}
		0, & \gamma<1\\
		\gamma - 1, & \gamma\geq 1
	\end{cases}.
\end{align}
We confirm the validity of Eq.~\eqref{eq_uE_Th} in Fig.~\ref{fig_num_var}(b). The scaling analysis to obtain $\alpha_{\mathrm{Th}}$ is shown in the inset of Fig.~\ref{fig_num_var}(b).

In terms of $\Ecal_{\mathrm{Th}}$, we corroborate in Fig.~\ref{fig_num_var}(a) that in the ergodic phase ($\gamma>1$), any two unfolded eigenvalues with spacing smaller than $\Ecal_{\mathrm{Th}}$ are correlated as in the GOE, while in the nonergodic phase ($\gamma<1$), they are uncorrelated. That is, for $\Delta_{\Ecal} < \Ecal_{\mathrm{Th}}$,
\begin{align}
	\Sigma^2(\Delta_\Ecal)\propto \begin{cases}
		\log(\Delta_\Ecal), &  \gamma>1\\ 
		\Delta_\Ecal, &  \gamma<1 .
	\end{cases}
\end{align}
We also verified that in the ergodic phase ($\gamma>1$), for a fixed value of $\gamma$, the level number variance $\Sigma^2(\Delta_\Ecal)$ deviates from the logarithmic behavior at larger values of $\Delta_\Ecal$ upon increasing $N$, that is, the Thouless energy grows with $N$ following Eq.~\eqref{eq_uE_Th}. Thus, for a fixed value of $\gamma>1$, the energy spectrum becomes more rigid as the matrix size increases. In contrast, the level number variance is system size independent at the critical point, $\gamma_c = 1$.

%=================================
%	Number variance and power-spectrum of noise
%=================================

A rather counterintuitive scenario for BRM is what happens for $\Delta_{\Ecal} > \Ecal_{\mathrm{Th}}$. As shown in Ref.~\cite{Erdos2015a}, above the Thouless energy, weak long-range correlations exist for BRM and follow the Altshuler-Shklovskii statistics~\cite{Altshuler1986}, 
\begin{equation}
	\Sigma^2(\Delta_\Ecal)\propto \sqrt{\Delta_\Ecal} ,
\end{equation}
irrespective of the localization properties of the eigenstates. This implies that even in the nonergodic phase with $0<\gamma<1$, any two unfolded eigenvalues with spacing larger than $\Ecal_{\mathrm{Th}}$ are weakly correlated. As a result, in the nonergodic phase, there is a transition from Poisson statistics ($\Sigma^2(\Delta_\Ecal) = \Delta_\Ecal$) to Altshuler-Shklovskii statistics ($\Sigma^2(\Delta_\Ecal) \propto \sqrt{\Delta_\Ecal}$) at $\Delta_\Ecal \approx \Ecal_{\mathrm{Th}}$, as analytically established in~\cite{Erdos2015a, Erdos2015b}. In the inset (i) of Fig.~\ref{fig_num_var}(a), we show the power-law behavior of the level number variance at large energy gaps along with a linear fit in the log-log scale. The slope of this fit gives the power-law growth exponent, $\alpha$, which is shown in the inset (ii) of Fig.~\ref{fig_num_var}(a). We observe that $\alpha \approx 1/2$, validating the square root behavior of the Altshuler–Shklovskii expression, that is, 
indeed $\Sigma^2(\Delta_\Ecal)\propto \sqrt{\Delta_\Ecal}$ for $\Delta_\Ecal \gg \Ecal_{\mathrm{Th}}$.

%============================================
\subsection{Power Spectrum}
%============================================

It has been shown in Ref.~\cite{Relano2002} that the energy spectrum fluctuations of quantum systems can be treated as a discrete time series and the power spectrum of this ``signal'' (that is, of these energy level fluctuations) distinguishes between regular (Brown noise) and chaotic ($1/f$ noise) phases. The focus of this analysis is the statistics 
\begin{align}
	\begin{split}
		\delta_n &= \sum_{i=1}^n (s_i - \mean{s}) = (\Ecal_{n+1} - n) -\Ecal_1,
	\end{split}
\end{align}
defined as the sum up to $n$ of the fluctuations of the spacings of unfolded levels, $s_i = \mathcal{E}_{i+1} - \mathcal{E}_{i}$, around the mean $\mean{s}$. The Fourier transform of $\delta_n$ is
\begin{equation}
	\hat{\delta}_k = \frac{1}{\sqrt{N}}\sum_{n}\delta_n e^{-\frac{2\pi i kn}{N}}  ,
\end{equation}
and the power spectrum is given by
\begin{align}
	\label{eq_P_Noise_def}
	P_k = |\hat{\delta}_k|^2.
\end{align}
In the chaotic phase,  $\mean{P_k}\propto k^{-1}$, and for integrable systems, $\mean{P_k}\propto k^{-2}$ \cite{Relano2002, Faleiro2004, Relano2008, Riser2017}.

In the case of BRMs with $0<\gamma<2$, the dependence of the properties of the energy correlations on the Thouless energy indicates that the behavior of $\mean{P_k}$ with $k$ should depend on $\gamma$ and whether $k$ is larger or smaller than a critical frequency $k_c$~\cite{Faleiro2004}. Since energy ($\delta_n$) is related to the power spectrum via a Fourier transform, the presence of energy correlations for $\Delta_\Ecal <\Ecal_{\mathrm{Th}}$ should get reflected in the behavior of $\mean{P_k}$ for $k>k_c$. On the other hand, the Altshuler-Shklovskii prediction of the existence of weak energy correlations for $ \Delta_\Ecal >\Ecal_{\mathrm{Th}}$ should get  reflected in the behavior of $\mean{P_k}$ for $k<k_c$. Consequently, we have
\begin{align}
	\mean{P_k} \propto \begin{cases}
		k^{-2}, & k > k_c,~ 0 < \gamma < 1\\
		k^{-1}, & k > k_c,~ 1 < \gamma < 2\\
		k^{-\frac{3}{2}}, & k < k_c,~ 0<\gamma<2
	\end{cases}.
	\label{eq:pwr_BRM}
\end{align}
The power spectrum has a homogeneous behavior only at three points: at $\gamma = 0$ (Poisson ensemble), where $\mean{P_k}\propto k^{-2}$ for all $k$, at $\gamma = 2$ (GOE), where $\mean{P_k}\propto k^{-1}$ for all $k$, and at $\gamma_c = 1$ (critical point), where $\mean{P_k}\propto k^{-\frac{3}{2}}$ for all $k$.

\begin{figure}[t]
	\centering
	\includegraphics[width=\columnwidth]{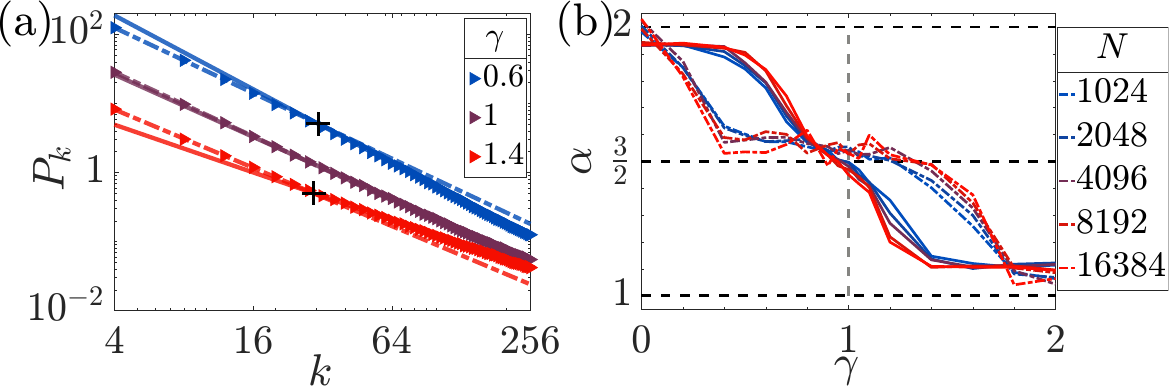}
	\caption{(a) Power-spectrum of noise for $N = 1024$ and different values of $\gamma$. The markers indicate numerical results and the dashed (solid) curves denote the linear fits of the low (high)-frequency behaviors in log-log scale. Crosses denote the critical frequency. The chosen energy windows are $\del{\Ecal-\frac{N}{8}, \Ecal+\frac{N}{8}}$ for $\Ecal$ in the middle 50\% of the spectrum.
		%-----------
		(b) Exponent $\alpha$ from $P_k \propto k^{-\alpha}$ as a function $\gamma$. Dashed line: power spectrum in the low-frequency ($k<k_c$) region; solid line: high-frequency ($k>k_c$).
	}
	\label{fig_power-spec}
\end{figure}

In Fig.~\ref{fig_power-spec}(a), we show numerical results via markers obtained for $\mean{P_k}$ vs $k$ for $\gamma$ in the nonergodic phase ($\gamma = 0.6$), ergodic phase ($\gamma = 1.4$), and at the critical point ($\gamma_c = 1$). We also show with dashed (solid) lines the linear fit of the low- (high-)frequency behavior in log-log scale. %, which clearly shows that $\mean{P_k} \propto k^{-\frac{3}{2}}$. The linear fit of the high-frequency behavior in log-log scale are shown via bold curves and 
The agreement between the numerics and the fitting curves validate Eq.~\eqref{eq:pwr_BRM}. The intersection of the low- and high-frequency behaviors happens at the critical frequency $k_c$, shown with crosses in  Fig.~\ref{fig_power-spec}(a).

%=================================
%	Number variance and power-spectrum of noise

In Fig.~\ref{fig_power-spec}(b), we show $\alpha$, from $P_k\propto k^{-\alpha}$, as a function of $k$ for various system sizes to demonstrate that as $N\rightarrow \infty$, the numerical results approach Eq.~\eqref{eq:pwr_BRM}. The dashed lines refer to results for $k > k_c$, and they approach $\alpha=2$ ($\alpha=1$) in the nonergodic (ergodic) phase. The solid lines are obtained for $k > k_c$ and show that $\alpha $ approaches the critical point $3/2$ as $N$ increases for all values of $\gamma$, except for the GOE and Poisson ensemble.

The presence of weak long-range energy correlation in the nonergodic phase, confirmed with the analysis of the level number variance in Fig.~\ref{fig_num_var}(a) and with the power spectrum in Figs.~\ref{fig_power-spec}(a)-(b), implies that two bulk eigenstates far apart in energy can weakly hybridize with each other, which contrasts with the exponentially localized eigenstates found in the case of Anderson localization. Such weak long-range energy correlations manifest as non-trivial temporal features in the SFF, as we discussed in Sec.~\ref{Sec:SFF_BRM}.

%============================================
\section{Conclusion}
%============================================
\label{Sec:Conclusion}

We investigated the level statistics of the banded random matrix (BRM) model in the nonergodic ($\gamma<1$) and ergodic ($\gamma>1$) phases, focusing on long-range energy correlations. This was done with a comprehensive analysis of the spectral form factor (SFF), level number variance, and power spectrum.

The analysis of the SFF involved deriving semi-analytical expressions for its entire temporal evolution and identifying its four characteristic timescales: Zeno time ($\tZ$), time for the onset of the ramp associated with correlated eigenvalues ($\tdip$), relaxation time ($\tR$), and Heisenberg time ($\tH$). The Zeno time determines the duration of the universal quadratic decay of the SFF and is independent of the phase. Beyond the Zeno time, we find that the SFF decays as a power-law $\propto t^{-3}$ for most values of $\gamma>0$ up to $\tdip$, when the ramp towards saturation begins. Notably, we verified that the correlation hole (dip-ramp-plateau structure) persists in the nonergodic phase. Furthermore, since these correlations decrease as $\gamma$ decreases below 1, $\tdip$ shrinks. In this phase, we observed that the relaxation time $\tR$ also decreases with $\gamma$, transitioning from $\tR \simeq \tH$ in the ergodic phase to $\tR = \tdip$ at $\gamma = 0$, where the correlation hole finally vanishes.

The energy correlations in the nonergodic phase of the BRM model differ from those in GOE full random matrices and disordered many-body quantum systems. In the latter case, as the disorder increases and the system approaches localization, long-range correlations vanish first, while remaining intermediate- and short-range correlations continue to follow the GOE. In the localized phase, all energy correlations of disordered many-body quantum systems should disappear. This contrasts with the BRM model, where weak long-range correlations persist in the nonergodic phase, despite the absence of short-range correlations, as confirmed by our SFF results. We also found signatures of weak long-range energy correlations in the power spectrum, $\mean{P_k}\propto k^{-\alpha}$. Above a threshold frequency $k_c$, the power-law exponent $\alpha$ distinguishes the nonergodic phase, ergodic phase, and critical point of the BRM. In addition, we showed that for $k<k_c$, $\alpha \to 3/2$ in any phase as the matrix size increases.

Our results are significant, as BRMs can model various physical systems. Verifying the properties identified here in experimental systems, particularly the persistence of energy correlations in the nonergodic phase,  would be a compelling direction for future research.

%=============================
%	---ACKNOWLEDGEMENT---
%=============================
\begin{acknowledgements}
	A.~K.~D. is supported by an INSPIRE Fellowship, DST, India and the Fulbright-Nehru grant no.~2879/FNDR/2023-2024. L.~F.~S. received supported from the National Science Foundation Center for Quantum Dynamics on Modular Quantum Devices (CQD-MQD) under Award Number 2124511.
\end{acknowledgements}

%======================================
%%%%% APPENDIX %%%%%
%======================================
\appendix
%
%=================================================
\section{Energy moments and density of states}
\label{sec_BRM_moms}
%=================================

We can compute the energy moments using Eq.~(\ref{eq_H_density}). The odd energy moments are zero for BRM, reflecting the symmetry of the density of states about $E = 0$. The second and fourth moments of energy are~\cite{Casati1991},
\begin{align}
	\label{eqA_E_2_4}
	\begin{split}
		\mean{E^2} &= \frac{F}{N},\quad \mean{E^4} = \frac{11 F + G - 5N}{2N}\\
	\end{split}
\end{align}
where $F = \frac{b}{2}(2N-b+1)$ and $G$  has a twofold nature due to corner and finite-size effects,
\begin{eqnarray}
	G &=& 2(b-1)\del{N(2b-3) - \dfrac{1}{3}b(5b-7)} \quad \text{for} \quad b\leq \dfrac{N}{2} ,\nonumber \\
	G &=& N(N-1)(N-2) \nonumber
	\\
	& &- \frac{2 (N-b)(N-b+1)(2N+b-5)}{3} \quad \text{for} \quad b > \dfrac{N}{2} \nonumber.
\end{eqnarray}

Based on Eq.~\eqref{eqA_E_2_4}, we define the shifted kurtosis, $K$ (normalized fourth order cumulant plus 1) as in Eq.~\eqref{eqA_dos_kurt}.
%\begin{align}
%	\label{eqA_dos_kurt}
%	K(b, N) = \frac{\mean{E^4}}{\mean{E^2}^2} - 2.
%\end{align}
$K(b, N)$ is 1 for a Gaussian distribution and 0 for the semi-circle. Equation~\eqref{eqA_dos_kurt} implies that $K(b, N) \approx \del{ \dfrac{1}{2b} + \dfrac{1}{3N} }$ for $b\ll N$ \cite{Casati1991}. For $b = cN$,
\begin{align}
	\label{eqA_dos_kurt_1}
	K(cN, N) = \begin{cases}
		\dfrac{2c(2-3c)}{3(c-2)^2}, & c \leq \dfrac{1}{2}\\
		\dfrac{2(c-1)^3(1 - 3c)}{3c^2(c-2)^2}, & c > \dfrac{1}{2}
	\end{cases}.
\end{align}
Equation~\eqref{eqA_dos_kurt_1} implies that $K(b, N)$ has a local maximum at $b = \dfrac{2N}{5}$ with a value $\dfrac{1}{12}$. Table~\eqref{tab_BRM_kurt} shows $K(b, N)$ for different values of $b$.
%============================================================================
%	Shifted Kurtosis at different limits
\begin{table}[b]
	\begin{tabular}{|c|c|c|}
		\hline
		$b$             & $K(b, N)$                                                     & $K(b, N\to\infty)$             \\ \hline
		1               & 1                                                             & 1                              \\ \hline
		$\sqrt{N}$      & $\dfrac{10N^{3/2} - 9N + 17\sqrt{N} - 6}{3(2N+1-\sqrt{N})^2}$ & 0                              \\ \hline
		$\dfrac{2N}{5}$ & $\dfrac{2(N+10)(8N+55)}{3(8N+5)^2}$                           & $\dfrac{1}{12} \approx 0.0833$ \\ \hline
		$\dfrac{N}{2}$  & $\dfrac{2(N+4)(N+14)}{3(3N+2)^2}$                             & $\dfrac{2}{27} \approx 0.0741$ \\ \hline
		$N$             & $\dfrac{3+N}{(1+N)^2}$                                        & 0                              \\ \hline
	\end{tabular}
	\caption{Values of the shifted kurtosis for BRM for different values of $b$.}
	\label{tab_BRM_kurt}
\end{table}
%=================================

%	DOS approximation
The density of states of a prominent random matrix ensemble, the Rosenzweig-Porter ensemble (RPE)~\cite{Rosenzweig1960, Kravtsov2015, Facoetti2016, Monthus2017, Bogomolny2018, Das2019, Soosten2019, Buijsman2022, Venturelli2023, DeTomasi2022, Das2023a, Sarkar2023}, has been obtained earlier~\cite{Bertuola2005}
\begin{align}
	\label{eqA_DOS_Bertuola}
	\rho(E; \alpha, a) = \frac{2}{\pi \alpha} \sqrt{\frac{a}{N}} \int_{0}^{\infty} dx \frac{\fbslj{1}{x}}{ x  e^{\frac{x^2}{4 N \alpha^2}} } \cos\del{\frac{\sqrt{a} E x}{\alpha \sqrt{N}}} ,
\end{align}
where $a$ is related to the variance of the off-diagonal matrix elements, $\fbslj{1}{x}$ is the Bessel function of the first kind of order 1 and $\alpha$ is the interpolation parameter. We assume that $\rho(E; \alpha, a)$ for RPE approximates the density of states for the BRM such that the energy moments of two random matrix ensembles can be equated to obtain
\begin{align}
	\label{eqA_E_mom_compare}
	\begin{split}
		\alpha^2 &= \frac{2(1-K + \sqrt{1-K})}{N K},\quad a = \frac{1 + \sqrt{1-K}}{2K\mean{E^2}} ,
	\end{split}
\end{align}
where the shifted Kurtosis $0 \leq K \leq 1$ and the second moment, $\mean{E^2}$, is given in Eq.~\eqref{eqA_E_2_4}. Upon scaling the energy axis as $E\to E/2\sqrt{\mean{E^2}}$, we get the following expression
\begin{align}
	\label{eqA_DOS_BRM_1}
	\begin{split}
		\rho(E, \kappa) &= \frac{2}{\pi}\sqrt{1+\frac{1}{\kappa}} \int_{0}^{\infty} dx \frac{ \fbslj{1}{x} \cos\del{ \sqrt{1+\frac{1}{\kappa}} E x} }{ x \exp\del{\frac{x^2}{8 \kappa}} } 
	\end{split}
\end{align}
where 
\begin{equation}
		\kappa \equiv \frac{1 - K + \sqrt{1-K}}{K}
\end{equation}
is Eq.~(\ref{eq_kappa}) in the main text and $\kappa = 0$ ($\kappa \to \infty$) for a Gaussian (semi-circle) DOS. Let $\omega = \sqrt{1+\frac{1}{\kappa}}E$, such that
\begin{align}
	\label{eqA_DOS_BRM_2}
	\rho_\omega\del{\kappa} = \frac{2}{\pi}\int_{0}^{\infty} dx \frac{\fbslj{1}{x} \cos\del{\omega x}}{ x\exp\del{\frac{x^2}{8 \kappa}} } .
\end{align}
Note that $\frac{\fbslj{1}{x}}{ x\exp\del{\frac{x^2}{8 \kappa}} }$ is an even function and $\cos\del{\omega x} = \frac{e^{i\omega x} + e^{-i\omega x}}{2}$. Let $\mathcal{F}[f]$ be the Fourier transform of $f(x)$,
\begin{align}
	\begin{split}
		\mathcal{F}[f] &\equiv \frac{1}{\sqrt{2\pi}} \int_{-\infty}^{\infty} f(x) e^{i\omega x}\\
		\Rightarrow \mathcal{F}\sbr{\frac{\fbslj{1}{x}}{x}} &= \sqrt{\frac{2(1-\omega^2)}{\pi}}\Theta(1-\omega^2)\\
		\mathcal{F}\sbr{ e^{-\frac{x^2}{8\kappa}} } &= 2\sqrt{\kappa} e^{-2\kappa \omega^2}
	\end{split}
\end{align}
Then, using the convolution theorem in Eq.~\eqref{eqA_DOS_BRM_2}, we get
\begin{align}
	\label{eqA_DOS_BRM_3}
	\begin{split}
		\rho_\omega(\kappa) &= \frac{\sqrt{8\kappa}}{\pi^{\frac{3}{2}}} \int_{-1}^{1} dx\: \sqrt{1 - x^2} e^{-2\kappa (\omega-x)^2}
		%\Rightarrow \rho(E, \kappa) &= \sqrt{\frac{8(\kappa+1)}{\pi^3}} e^{-2(\kappa+1)E^2}\\
		% &\times \int_{-1}^{1} dx \sqrt{1-x^2} e^{-2\kappa x^2 + 4\sqrt{\kappa^2+\kappa} Ex}.
	\end{split}
\end{align}
such that the density of states of BRM is [Eq.~(\ref{eq_DOS_BRM}) in the main text]
\begin{align}
	\label{eq_DOS_BRM_APP}
	\begin{split}
		\rho(E, \kappa) &= \frac{ \sqrt{\frac{8(\kappa+1)}{\pi^3}} }{e^{2(\kappa+1)E^2}} \int_{-1}^{1} dx \frac{\sqrt{1-x^2}}{ e^{2\kappa x^2 + 4\sqrt{\kappa^2+\kappa} Ex} }.
	\end{split}
\end{align}
The above expression allows for a closed form at $E = 0$,
\begin{align}
	\label{eq_DOS_E0}
	\rho(0, \kappa) = \sqrt{\frac{2(\kappa+1)}{\pi}} \frac{\fbsli{0}{\kappa} + \fbsli{1}{\kappa}}{e^\kappa}
\end{align}
where $\fbsli{m}{x}$ is the modified Bessel function of the first kind of order $m$.  For $\kappa = 0$, Eq.~\eqref{eq_DOS_BRM_APP} yields the Gaussian distribution $\sqrt{\frac{2}{\pi}} e^{-2E^2}$, valid for the Poisson ensemble ($\gamma = 0$). For $\kappa\to \infty$, $\omega \approx E$ and we get
\begin{align}
	\begin{split}
		\rho(E, \infty) &\approx \frac{2}{\pi} \int_{-1}^{1} dx \sqrt{1-x^2} \times \sqrt{\frac{2\kappa}{\pi}} \exp\del{-\frac{(x-E)^2}{2\frac{1}{4\kappa}}}\\
		% &\approx \frac{2}{\pi} \int_{-1}^{1} dx \sqrt{1-x^2} \delta\del{x-E}\\
		&\approx \frac{2}{\pi} \int_{-\infty}^{\infty} dx \sqrt{1-x^2}\Theta(1-x^2) \delta\del{x-E}\\
		&= \frac{2}{\pi}\sqrt{1-E^2}, \quad |E|\leq 1 ,
	\end{split}
\end{align}
where $\Theta(x)$ is the Heaviside step function and we approximated the narrow Gaussian distribution with the Dirac delta distribution, $\delta(x)$. Thus from Eq.~\eqref{eqA_DOS_BRM_3}, we recover the semi-circle law for $\kappa\to\infty$, valid for the GOE ($\gamma = 2$).

%=================================

%======================================
%%%%% REFERENCES %%%%%
%======================================
\bibliography{ref_BRM}

\end{document}